\documentclass[runningheads]{svmult}

\usepackage{makeidx}
\usepackage{graphicx}
\usepackage{subeqnar}
\usepackage{multicol}
\usepackage{physmubb}
\usepackage{amssymb}
\usepackage{amsmath}
\makeindex 

\begin{document}

\title*{Velocity fluctuations in cooling granular gases.}

\toctitle{Cooling granular gases: the role

\protect\newline of correlations in the velocity field}

\titlerunning{Cooling granular gases}

\author{Andrea Baldassarri~\inst{1}
\and Umberto Marini Bettolo Marconi~\inst{2}
\and Andrea Puglisi~\inst{3}}

\institute{INFM Udr Roma 1, University of Rome ``La Sapienza'', piazzale Aldo Moro 2, I-00185, Roma, Italy \and Dipartimento di Matematica e Fisica and INFM Udr Camerino, University of Camerino, Via Madonna delle Carceri
I-62032 Camerino, Italy \and INFM Center for Statistical Mechanics and Complexity, University ``La Sapienza'', piazzale
Aldo Moro 2, I-00185 Roma, Italy}

\authorrunning{Andrea Baldassarri et al.}

\maketitle

\begin{abstract}
We study the formation and the dynamics of correlations in the
velocity field for 1D and 2D cooling granular gases with the
assumption of negligible density fluctuations (``Homogeneous
Velocity-correlated Cooling State'', HVCS). It is shown that the
predictions of mean field models fail when velocity fluctuations
become important. The study of correlations is done by means of
molecular dynamics and introducing an Inelastic Lattice Maxwell
Models. This lattice model is able to reproduce all the properties of
the Homogeneous Cooling State and several features of the
HVCS. Moreover it allows very precise measurements of structure functions
and other crucial statistical indicators. The study suggests that both
the 1D and the 2D dynamics of the velocity field are compatible with a
diffusive dynamics at large scale with a more complex behavior at
small scale. In 2D the issue of scale separation, which is of interest
in the context of kinetic theories, is addressed.
\end{abstract}

\section{Introduction}

The inelastic hard spheres model~\cite{poschel00} without energy
input, initially prepared in a homogeneous state, exhibits, after a
rapid transient, a regime characterized by homogeneous density and a
probability distribution of velocities that depends on time only
through the total kinetic energy (global granular temperature
$T_g(t)$), i.e. a scaling velocity distribution. This is the so called
{\em Homogeneous Cooling State} (HCS) or Haff regime~\cite{haff83}. It
has been shown by several
authors~\cite{goldhirsch93,mcnamara92,deltour97,noije97} that this
state is unstable with respect to shear and clustering instabilities:
structures can form that seem to minimize dissipation, mainly in the
form of velocity vortices and high density clusters. These
instabilities grow on different space and time scales, so that one can
investigate them separately. Several theories have been proposed to
take into account the emergence of structures in granular gases. Some
of these are more fundamental, because derive the correlation functions directly from
the kinetic equations~\cite{noije98}; others, that assume the validity
of the hydrodynamic description, deserve the name of mesoscopic
theories~\cite{noije00}; others are phenomenological theories that
suggest analogies with Burgers equation~\cite{ben-naim99,ben-naim02}
or spinodal decomposition models~\cite{wakou02} or mode coupling
theories~\cite{brito98}. Some of these theories can describe the behavior
of the cooling granular gas far deeply into the correlated regime,
giving predictions for the asymptotic decay of energy. In this paper,
after a brief summary of the results of these theories, we show how
the departure from the HCS can be well modeled by a class of models
obtained placing on a lattice the original homogeneous Inelastic
Maxwell Model. These models have the disadvantages of being conceived
under the assumption of negligible density fluctuations and therefore
can be useful only in the description of the first instability,
i.e. the growth of velocity fluctuations. In section 2 we review the
HCS, its instabilities and the existing theories. In section 3 we
briefly discuss the Inelastic Maxwell
Model~\cite{baldassarri_epl,baldassarri_trieste,baldassarri_mmm},
which is a starting point for the introduction of the Inelastic
Lattice Maxwell Models (ILMM). In section 4 and in section 5 the
analysis and results of the ILMM in one and two dimensions are
reviewed~\cite{baldassarri_trieste,baldassarri_pre}. Finally in section 6
conclusions are drawn.

\section{Instabilities of the Homogeneous Cooling State}

A cooling granular gas in $d$ dimensions is defined as an ensemble of
$N$ grains, i.e. hard objects (rods if $d=1$, disks if $d=2$, spheres
if $d=3$) of linear size (diameter) $\sigma$, placed in a volume $V$
with periodic boundary conditions. The grains evolve freely and
interact with each other through instantaneous binary inelastic
collisions. The rule that gives the velocities after the collision as
functions of the velocities before the collision is the definition of
the particular granular gas model. In this case we use the model with
constant restitution coefficient without rotational degrees of freedom
and set the mass of the grains $m=1$. Other models have been
considered in the literature~\cite{poschel00}. The collision rule between a particle
with velocity $\mathbf{v}$ and one with velocity $\mathbf{v}^*$ for
this model is:

\begin{equation}
\mathbf{v}'=\mathbf{v}-\frac{1+r}{2}[(\mathbf{v}-\mathbf{v}^*)\cdot \hat{\mathbf{n}}]\hat{\mathbf{n}}
\end{equation}
where the primed velocity is the post-collisional one,
$\hat{\mathbf{n}}=(\mathbf{r}-\mathbf{r}^*)/|\mathbf{r}-\mathbf{r}^*|$
is the unit vector in the direction joining the centers of colliding
particles, and $r$ is the restitution coefficient and takes values
between $0$ and $1$. When $r=1$ the collision is elastic.

Usually (in numerical or real experiments) granular gases are prepared
in a homogeneous situation: uniform-random positions of grains,
Gaussian or uniform-random initial velocity with no preferred
direction. It happens that, if the system is large enough or the
inelasticity large enough, the imposed homogeneity is broken after a
certain time. The more accepted scenario is a two time homogeneity
breaking: at a time $t_s$ the velocity field becomes unstable to the
formation of shear bands, then at a time $t_c>t_s$ the density field
becomes unstable toward the formation of high density clusters. 

\subsection{The homogeneous cooling state}

A granular gas evolving from a homogeneously random state loses memory
of its initial condition after a time of the order of one collision
per particle and rapidly enters the Homogeneous Cooling
Regime. This regime is expected to be well described by the granular
Boltzmann Equation (see for example~\cite{noije98}). This means that in
this regime the total probability density function (p.d.f.) at
collision is factorisable in one-particle p.d.f.'s:
$p_N(\mathbf{r}_1...\mathbf{r}_N,\mathbf{v}_1...\mathbf{v}_N,t)=\prod_{i=1}^N
p_1(\mathbf{r}_i,\mathbf{v}_i,t)$.

The kinetic definition of HCS is given by the homogeneity ansatz plus
the scaling ansatz for the one particle distribution function:

\begin{equation}
\label{APscaling_pdf}
p_1(\mathbf{r},\mathbf{v},t)=\frac{1}{V}P(\mathbf{v},t)=\frac{1}{Vv_0^d(t)}\tilde{P}(\mathbf{c})
\end{equation}
where $\mathbf{c}=\mathbf{v}/v_0(t)$ and
$v_0(t)$ is the thermal velocity defined by $T(t)=v_0^2(t)/d$ with
$T(t)$ the granular temperature; here we have assumed that $\int d\mathbf{v}
\int d\mathbf{r}p_1=1$ and also that $\int d\mathbf{v}P=1$.  If the Eq. \eqref{APscaling_pdf} is
inserted in the Boltzmann Equation an equation for the temperature is
obtained:

\begin{equation}
\label{APtemperature_equation}
\frac{dT}{dt}=-2\omega\gamma T
\end{equation}

where $\omega$ is the time dependent collision frequency, while
$\gamma$ is the time independent cooling rate. These two
functions can be approximated, using the Maxwellian approximation $\tilde{P}
\approx (\frac{2\pi}{2})^{-d/2}\exp(-\frac{dc^2}{2})$, by $\omega_0$ and $\gamma_0$:

\begin{subequations}
\begin{align}
\omega_0 &\propto \sqrt{T} \\
\gamma_0 &=\frac{1-r^2}{2d}.  \label{APgamma_zero}
\end{align}
\end{subequations}
In this case the solution of the temperature equation \eqref{APtemperature_equation}
reads:

\begin{equation}
\label{APcooling_temperature}
T(t)=\frac{T(0)}{(1+\gamma_0 t/t_0)^2}=T(0)\exp(-2\gamma_0 \tau)
\end{equation}

where $t_0=1/\omega(T(0))$ is the mean free time at the initial temperature $T(0)$ and 

\begin{equation}
\tau=\frac{1}{\gamma_0}\ln(1+\gamma_0+t/t_0)
\end{equation}
is the cumulated collision number obtained from the definition
$d\tau=\omega(T(t))dt$. Eq. \eqref{APcooling_temperature} is known as
Haff's law~\cite{haff83}. 

Corrections to the constants appearing in
eq.~\eqref{APcooling_temperature} stem from a more careful consideration
of the HCS: when the volume fraction is non negligible the
Enskog-Boltzmann equation should be employed instead of Boltzmann
equation. This is identical to the Boltzmann equation but for a
multiplicative constant in the collision integral that takes into
account static density correlations due to the fact that the gas is
not perfectly dilute. Very recently it has been shown~\cite{poschel_private}
that there are also small velocity correlations which must be
considered and that modify the molecular chaos hypothesis: with this
correlations the law $T \sim t^{-2}$ is still valid but more precise
corrections to the constant $\gamma_0/t_0$ appear.  

The Haff law can be also derived in the framework of granular
hydrodynamics~\cite{brey98,noije97}: in the HCS hydrodynamics can be
considered valid as a consequence of homogeneity of the density and
velocity field and of the slow temperature decay. The evolution of
the (space uniform) temperature field is exactly the same as in
eq.~\eqref{APcooling_temperature}.

\subsection{Instabilities of the homogeneous cooling state}

Kinetic arguments, linear stability analysis of hydrodynamic equations
and numerical simulations lead to the evidence that the HCS is
unstable. In particular, it has been shown that large scale
fluctuations of velocity and density can grow exponentially breaking
the homogeneity of the system. The fact that the fluctuations arise at
large scale means that, if the system is small enough, the HCS lasts
forever. Moreover the minimum scale of formation of fluctuations
depends upon inelasticities and is smaller at larger inelasticities:
this means that for a quasi-elastic gas it is necessary to take a very
large volume to observe the breaking of homogeneity.

\begin{figure}[tbp]
\centerline {\includegraphics[clip=true,width=.7\textwidth,keepaspectratio]{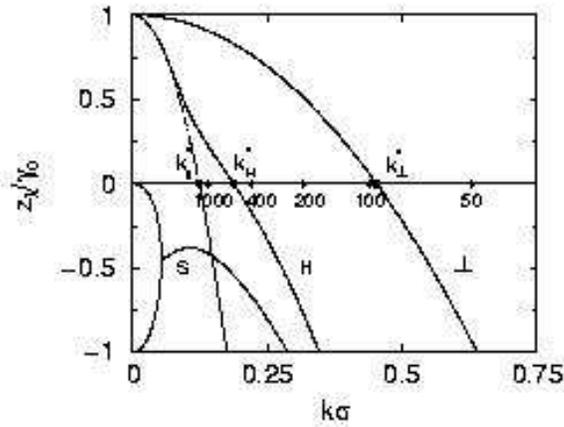}}
\caption
{Growth rates $\zeta_\lambda/\gamma_0$ for shear ($\lambda=\perp$),
heat ($\lambda=H$) and sound ($\lambda=\pm$) modes versus $k\sigma$
for inelastic hard disks with $r=0.9$ at a packing fraction
$\phi=0.4$. The dashed line indicate the imaginary parts of the sound
modes. Here $\sigma$ is the diameter of a particle. (From Orza et
al.~\cite{orza97})\label{APfig_disp_rel}}
\end{figure}

The linear stability analysis of hydrodynamics gives a good
description of the instability of fluctuations. The procedure is
applied to the fields rescaled with respect to the Homogeneous Cooling
State, i.e. $\tilde{n}=n(\mathbf{r},t)/n$ (being $n$ the average density),
$\tilde{\mathbf{u}}(\mathbf{r},t)=
\mathbf{u}(\mathbf{r},t)/\sqrt{T(t)}$ and $\tilde{T}(\mathbf{r},t)=T(\mathbf{r},t)/T(t)$,
and therefore any instability or emerging structure is not absolute
but relative to this decaying reference state. Moreover the
instabilities are studied in $\mathbf{k}$ space (the space of Fourier
modes) and, when $d>1$, the velocity field is decomposed in a parallel
component $\tilde{\mathbf{u}}_{\parallel}(\mathbf{k},t)$ and in $d-1$
orthogonal components $\tilde{\mathbf{u}}_{\perp}(\mathbf{k},t)$
with respect to the vector $\mathbf{k}$. In
figure~\ref{APfig_disp_rel} we report the results of such an analysis,
taken from the literature~\cite{orza97}.  This analysis shows that the
evolution of fluctuations of normal velocity components (shear modes,
$\tilde{\mathbf{u}}_\perp$) are not coupled with any other fluctuating
component. At the same time, the remaining components are coupled
together.  The rate of decay/growth for the shear mode reads
$\zeta_\perp(k)=\gamma_0(1-k^2\xi_\perp^2)$ where $\xi_\perp$ depends
on the transport coefficients appearing in the hydrodynamics. We refer
to~\cite{noije00} for detailed calculations of this correlation
length. At low values of $k$ (in the so-called ``dissipative range'')
also the heat mode is ``pure'', as it is given by the longitudinal
velocity mode $\tilde{\mathbf{u}}_\parallel$ only, with eigenvalue
$\zeta_H(k) \simeq \gamma_0(1-\xi_\parallel^2k^2)$; in this range the
sound modes are combination of density and temperature fluctuations.
The most important result of this analysis is that $\zeta_\perp(k)$
and $\zeta_H(k)$ are {\em positive} below the threshold values
$k_\perp^*=1/\xi_\perp\sim \sqrt{\epsilon}$ and $k_H^* \simeq
1/\xi_\parallel\sim \epsilon$ respectively, indicating two linearly
unstable modes with exponential (in $\tau$, i.e. the time measured by
the cumulated number of collisions per particle) growth rates. Here we
have used the notation $\epsilon=1-r^2\equiv 2d\gamma_0$.

The shear and heat instabilities are well separated at low
inelasticity (i.e. low $\epsilon$), as $k_\perp^* \sim
\sqrt{\epsilon}$ while $k_H^* \sim
\epsilon$, so that $k_\perp^* \gg k_H^*$. It is also important to note that 
the linear total size $L$ of the system can suppress the various
instability, because the minimum wave number $k_{min}=2\pi/L$ can be larger
than $k_H^*$ or even than $k_\perp^*$. 

When fluctuations grow, structures emerge. In molecular dynamics
simulations these structure appear as shocks in $d=1$, or vortices and
clusters in $d=2$. In $d>2$, a detailed analysis of structure factors
(from fluctuating hydrodynamics~\cite{noije00} or ring kinetic
theory~\cite{noije98}) has established that all the main
correlation lengths (which indicate the typical size of structures
{\em in the rescaled fields}) should grow as $\sim \tau^{1/2}$. In
particular this is expected for the size of vortices $L_v \sim
\tau^{1/2}$ and the size of clusters $L_{cl} \sim
(\tau/\epsilon)^{1/2}$: this again shows that, at small inelasticities
$\epsilon$, the size of clusters grows more slowly than that of
vortices.

\subsection{Conjectures}

After a brief transient during which the instabilities grow
exponentially in $\tau$ with the rates given above, the description
suggested by linear stability analysis ceases to be
meaningful. Molecular dynamics (MD), in the absence of experiments (the
cooling granular gas being a reference model and not a real system),
are the more reliable tools, but they are very cpu-time consuming: to
obtain a decent statistics of the asymptotic dynamics of a granular
gas, large enough to observe instabilities, MD simulations must be
carried on for days. To our knowledge a definitive clear picture of
the asymptotics of this model is still lacking (with month-lasting
simulations Isobe~\cite{isobe_private} has obtained what he called the
``final state'', which resembles 2D turbulence, but this was done in a
quasi-elastic limit). 

On the other side there are effective models that can grasp the
essence of the phenomena with particular assumptions. Here we rapidly
review some of these:

\begin{itemize}

\item
Brito and Ernst have shown by means of a mode-coupling~\cite{brito98}
theory that the asymptotic energy decay should follow a diffusive form
$\sim \tau^{-d/2}$.

\item
Wakou et al.~\cite{wakou02} have demonstrated that the evolution of
the flow field of a granular fluid, neglecting the convective term
$\mathbf{u}\cdot \boldsymbol{\nabla}\mathbf{u}$, can be cast in the
form of a Time Dependent Ginzburg-Landau equation for a non-conserved
order parameter.  The energy functional has a {\em continuous} set of
degenerate minima, having the shape of a Mexican hat. When the
convective term is added, only subsets of admissible solutions are
selected out of this infinite set of minima. In two dimensions only
{\em two} distinct minima survive: therefore the $d=2$ cooling
granular gas, during the formation of instabilities, greatly resembles
the spinodal decomposition for a non-conserved order parameter (Model
A universality class, in the Hohenberg-Halperin
classification~\cite{hohenberg77}).

\item
Ben-Naim et al.~\cite{ben-naim99} have studied the cooling granular
gas in $d=1$, discovering that, after the homogeneous (Haff) phase, it
asymptotically becomes independent of the value of inelasticity
$\epsilon$ and maps onto the sticky gas model. For such a sticky gas
the asymptotic temperature decay is proportional to $t^{-2/3}$, when
density clusters and velocity shocks form. The velocity field of the
sticky gas~\cite{carnevale90} is described by the inviscid limit of
the Burgers equation~\cite{shandarin89}.  The relation to the Burgers
equation is useful to establish an estimate of the tails of the
asymptotic velocity distribution $P(v,t)
\sim t^{1/3}\Phi(vt^{1/3})$ (independent of $r$), which reads $\Phi(z)
\sim \exp(-|z|^3)$. The MD simulations have revealed only slight deviations
from the Gaussian, and the authors have imputed the discrepancy from
the expected behavior to the smallness of the constant in front of
$|z|^3$. More recently it has been conjectured~\cite{ben-naim02} that
the Burgers equation describes the flow velocity
$\mathbf{u}(\mathbf{r},t)$ of a cooling granular gas for arbitrary
values of the inelasticity $\epsilon$ in generic $d$ dimensions. This
conjecture implies that the asymptotic behavior of the cooling
inelastic gas is independent of $\epsilon$ (as it always falls in the
universality class of the sticky gas) and that the upper critical
dimension for the disappearance of the inelastic collapse is
$d_c=4$. However MD simulations in $d>4$ seem to show that the shear
instability and the inelastic collapse do not disappear~\cite{barrat00}.

\end{itemize}

\section{A starting point: the homogeneous Inelastic Maxwell Model}

To better illustrate the analysis of the Inelastic Lattice Maxwell
Models (ILMM) we start with a brief description of the homogeneous
Inelastic Maxwell Model (IMM). The definition of the model is given in
terms of its Boltzmann Equation, which, for scalar velocities ($d=1$),
reads:

\begin{equation}
\label{APpseudo_Maxwell_eq_hom1d}
\frac{\partial P(v,\tau)}{\partial \tau} +P(v,\tau)=  \frac{1}{1-\gamma^*}\int duP(u,\tau)P\left(\frac{v-\gamma^* u}{1-\gamma^*},\tau\right)
\end{equation}
with $\gamma^*=(1-r)/2$. The IMM can be introduced in different ways:
(a) its Boltzmann equation can be regarded as the master equation of a
simple stochastic process inspired to a model put forward by
S.Ulam~\cite{ulam80}: the process consists in the evolution of an
ensemble of $N$ velocities, at each time step (corresponding to
$\Delta \tau=2/N$, so that $\tau$ increases of $1$ after averagely one
collision per particle) two velocities are chosen at random with uniform
distribution and are changed according to the inelastic collision
rule; (b) the associated Boltzmann equation can be considered the inelastic
generalization of the equation for elastic Maxwell
Molecules~\cite{maxwell1867,ernst81}; (c) its Boltzmann equation can
be derived from the ordinary Boltzmann equation for inelastic hard
spheres assuming that the term $|v-v'|$ in the collision integral can
be approximated by $\sqrt{T}$, as suggested by Bobylev et
al.~\cite{bobylev00}.

The recent surge of interest for this model has been triggered by the
discovery of an exact
solution~\cite{baldassarri_epl,baldassarri_trieste,baldassarri_mmm}
for the $d=1$ case (eq.~\eqref{APpseudo_Maxwell_eq_hom1d}) and the
appearance of computable power law tails in the solutions of $d>1$
cases.~\cite{ernst02,ben-naim02b}. In a previous paper Ben-Naim and
Krapivsky~\cite{ben-naim00} showed that
eq.~\eqref{APpseudo_Maxwell_eq_hom1d} could not have a scaling
solution with finite moments: in fact, a scaling form

\begin{equation}\label{scalingsol}
P(v,\tau)\to \frac{1}{v_0(\tau)}f(\frac{v}{v_0(\tau)})
\end{equation}
with $v_0^2(\tau)=\int v^2 P(v,\tau)dv=E(\tau) \sim \exp(-\tau a_2)$, imposes a temporal dependence
of the higher moments (if finite):

\begin{subequations}
\label{APscaling-moments}
\begin{align}
\langle v^{2m}\rangle &= v_0(\tau)^{2m} \mu_{2m}\\
\text{ where }\; \mu_{2m}&=\int dy f(y) y^{2m}
\end{align}
\end{subequations}
does not depend on time. But they found that 

\begin{equation}
\lim_{\tau\to\infty} \frac{\langle v^{2m}(\tau)\rangle}{\left(\langle
v^2(\tau)_\rangle\right)^m}=\infty.
\end{equation}

This result does not exclude the existence of a scaling
solution~\eqref{scalingsol} but requires a solution with diverging
moments of order $2m \geq 4$, i.e. very large algebraic tails. In this
case standard methods~\cite{ernst81} based on Fourier transform fails
due to the singularity in $k=0$ corresponding to such algebraic tails.

\begin{figure}[tbp]
\centerline {\includegraphics[clip=true,width=.7\textwidth,keepaspectratio]{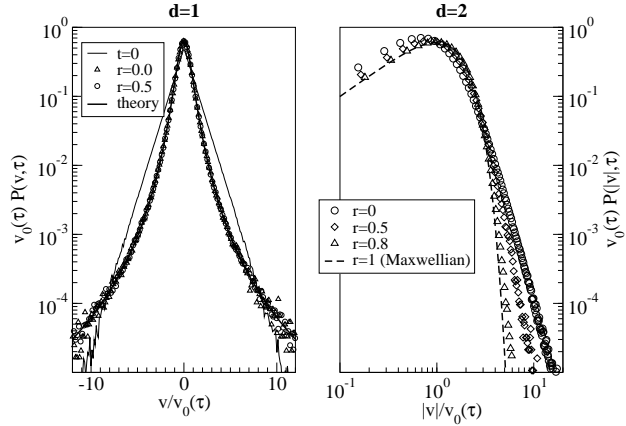}}
\caption
{Asymptotic velocity distributions $P(v,\tau)$ versus $v/v_0(\tau)$
for different values of $r$ from the simulation of the Inelastic
Maxwell Model in 1D (left) and 2D (right). In 1D the asymptotic
distribution is independent of $r$ and collapse to the
Eq.~\eqref{APasymptotic_solution}. In 1D, the chosen initial
distribution (exponential) is drawn (same result with uniform and
Gaussian initial distribution).  In 2D the distributions still present
power-law tails, but the power depends upon $r$: for $r\to 1$ the pdf
tends to a Maxwell distribution. Data refers to more than $N=10^6$
particles.\label{APfig_maxwell}}
\end{figure}

Remarkably, a direct inspection shows that the following velocity distribution

\begin{equation}
\label{APasymptotic_solution}
P(v,\tau)=\frac{2}
{\pi v_0(\tau)\left[1+\left(\frac{v}{v_0(\tau)}\right)^2\right]^2}
\end{equation}
is indeed a solution of the non-linear Boltzmann
equation~\eqref{APpseudo_Maxwell_eq_hom1d}, for every value of
$r$~\cite{baldassarri_epl,baldassarri_trieste,baldassarri_mmm}. The
solution~\eqref{APasymptotic_solution} has the property that the moments of order $2m
\geq 4$ diverge, and its Fourier transform has a singularity of the type $|k|k^2$ in $k=0$.

We believe that eq.~\eqref{APasymptotic_solution} represents the
asymptotic solution for a large class of initial distributions, for
the following arguments:
\begin{itemize}
\item
as shown by Ben Naim and Krapivsky~\cite{ben-naim00}, the dynamics of
the moments for a generic starting distribution can be computed,
giving the following limit $\lim_{t \to
\infty} \langle v^{2m}(t)\rangle / \langle v^2(t)\rangle ^m =\infty$
for $m>1$;
\item
secondly we performed numerical simulations of the BK model,
collecting evidence of the convergence to the
solution~\eqref{APasymptotic_solution} for several starting velocity
distributions, namely uniform, exponential (see figure~\ref{APfig_maxwell},
left frame) or Gaussian.
\end{itemize}

For higher dimensions, the problem of a scaling solution has been
recently addressed in~\cite{ben-naim02,ernst02}. Assuming a
solution with large algebraic tails, it is possible to compute the
asymptotic exponent as a solution of a transcendental equation.
Such equation depends explicitly on the restitution coefficient,
indicating that the tail exponent varies with $r$ (and correctly
diverges for $r\to 1$, corresponding to the recovering of the Maxwell
distribution for the elastic case).

\begin{figure}[tbp]
\centerline {\includegraphics[clip=true,width=.7\textwidth,keepaspectratio]{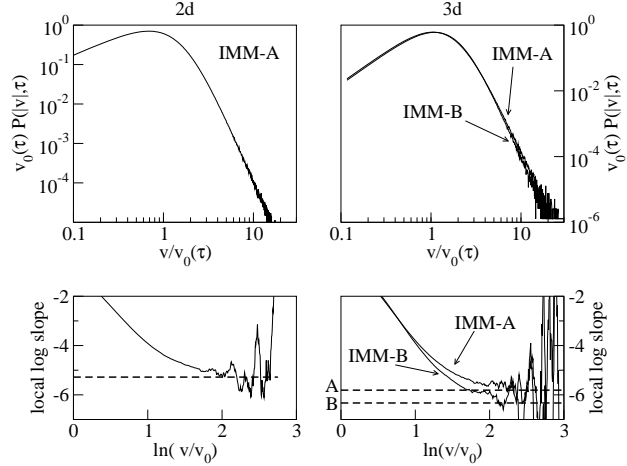}}
\caption
{Asymptotic velocity distributions $P(|v|,\tau)$ versus $|v|/v_0(\tau)$
for $r=0$ from the simulation of the Inelastic
Maxwell Model in 2D (left) and 3D (right). In 2D the asymptotic
distribution tail converges to the exactly computed exponent (model
IMM-A). In 3D the convergence is slower, but the difference of the predicted
exponent for model IMM-A and IMM-B can be appreciated. In the bottom
panels the convergences of the local logarithmic slope is put in
evidence. (For the precise definition of the two models
see~\protect{\cite{ernst02}}.) Data refers to more than $N=10^6$
particles.\label{APfig_tails}}
\end{figure}

A numerical observation of such algebraic tails in $2$ dimensions are
shown in right frame of figure~\ref{APfig_maxwell}, where a
dependence on $r$ is put in evidence.
A more precise comparison of the measured exponent and the theoretical
predictions are shown in figure~\ref{APfig_tails}.
In this figure we show the results of extensive numerical simulations
for the completely inelastic case ($r=0$) of several models, in two and
three dimensions.

In fact, as noted in~\cite{ernst02}, the tail exponent is in general
dependent on the details of the collisions, as for instance on the
distribution of the impact parameter. Two different choices have
been studied in $3$ dimensions, corresponding to the models denoted
IMM-A and IMM-B in reference~\cite{ernst02}: in the first model grazing
collisions are favored with respect to the second one.
It is interesting to notice that the analytical and numerical study
confirm the existence of a scaling solution with algebraic tails,
irrespectively of the collision details, but that the precise value of
the exponent is not universal.

The relevance of such a scaling solution for $d>1$ has been addressed
by Bobylev {\em et al.}~\cite{bobylev03}, who have proved the
existence and the uniqueness of such an asymptotic solution (giving
details on the basin of attraction).

Interesting features emerge when the inelastic Maxwell model is
extended to treat grains having different physical properties, such as
unequal masses, different restitution coefficients, radii, etc.  The
study of a binary Maxwell mixture with scalar velocities showed few
unexpected properties~\cite{bettolo02a,bettolo02b}:

\begin{enumerate}

\item
the kinetic temperatures of the two species are different, i.e.
there is no energy equipartition, but the the ratio $T_1/T_2$
reaches asymptotically a constant value. Such a feature agrees with the 
results obtained in the framework of the Boltzmann-Enskog transport equation
by Garz\`o and Dufty \cite{garzo99}.

\item
The velocity distributions in general differ in shape.

\item
The velocity distributions have power law tails. These exponents
can be calculated analytically, by a suitable generalization
of a technique devised by Ernst and Brito \cite{ernst02}.
The power law decay of cooling mixtures has been shown to vary between
$2$ and $\infty$.

\end{enumerate}

\section{The one-dimensional gas}

The solution of the IMM given in~\eqref{APasymptotic_solution} has
nothing to do with the real dynamics of a cooling granular gas in the
homogeneous regime, where no algebraic tails appear. Instead, a good
agreement (in the HCS) is found between the solution of Boltzmann
equation~\cite{benedetto97} and Molecular
Dynamics~\cite{barrat02}. However we will show in the following that
the Inelastic Maxwell Model with the addiction of a special topology
(the 1d lattice) is able to reproduce the homogeneous regime and some
essential features of the first instability (enhancement of velocity
fluctuations).

\subsection{Molecular Dynamics}

An MD simulation of $N$ hard rods on a ring of length $L$ shows that,
besides hydrodynamic instabilities discussed above, also a
kinetic singularity appears that prevents long simulation runs, the
so-called ``inelastic collapse''~\cite{mcnamara92,mcnamara94}. The
inelastic collapse is the divergence of the collision rate of a small
set of grains, analogous to the divergence of the collision rate of a ball
left bouncing on the floor. This means that a group of particles will
experience an infinite number of collisions in a finite time. To avoid
such a problem, many authors~\cite{ben-naim99,luding98} have proposed the
introduction of a velocity cut-off $\delta$ such that, whenever two
particles collide with a relative velocity smaller, in absolute value,
than $\delta$, the collision is elastic. Taking $\delta^2 \ll
T_{min}$, being $T_{min}$ the minimum granular temperature that one
expects to observe, one can be sure that the choice of $\delta$ will
not influence the main results of the simulation.
Performing an efficient Event Driven simulation scheme (based on a
tree ordering of the event times) it is possible to study easily 
very large systems, with $N \sim 10^6$ for thousands of collisions per particle.

\begin{figure}[tbp]
\centerline {\includegraphics[clip=true,width=.7\textwidth,keepaspectratio]{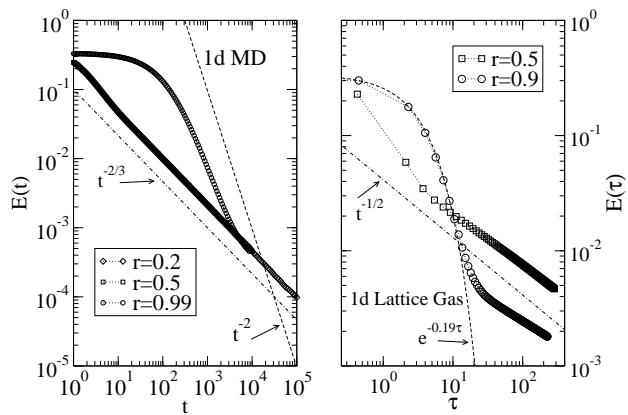}}
\caption
{Time behavior of kinetic energy for the Inelastic Hard Rods (left)
and for the Inelastic Lattice Maxwell Model (right). The homogeneous
Haff stage is evident only for quasi-elastic system, whereas the more
inelastic ones enter almost immediately in the correlated regime. Note
the different time units used ($t$ is the physical time, defined only
in MD simulations, while $\tau$ is the cumulated number of collisions
per particle). The Haff law $E \sim t^{-2}
\sim \exp(-\gamma_0\tau)$ is verified for both systems. The correlated regime
presents a behavior $t^{-2/3}$ independent of $r$ for the Hard Rods,
while appears diffusive (in collision units) $\tau^{-1/2}$ and
$r$-dependent for the Lattice Model. Data refer to $N=10^6$ particles
(both models).\label{APfig_energia_1d} }
\end{figure}

\begin{figure}[tbp]
\centerline {\includegraphics[clip=true,width=.7\textwidth,keepaspectratio]{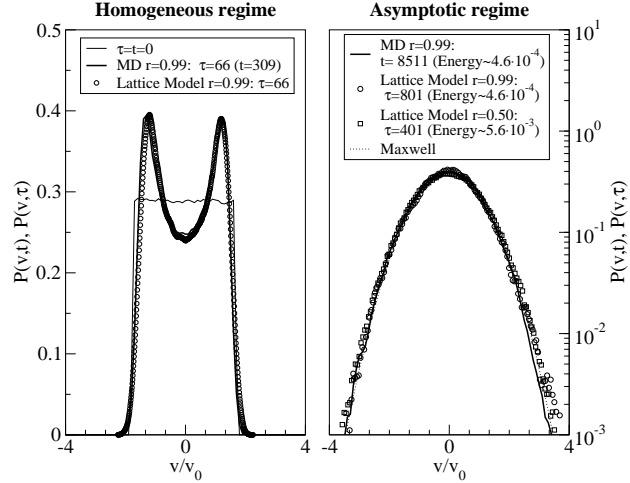}}
\caption{
Rescaled velocity distributions for the 1D MD and in the 1D ILMM,
during the homogeneous (left) and the inhomogeneous phase (right). In
the left frame is also shown the initial distribution (both
models). The distributions refer to systems having the same
energy. Data refer to  $N=10^6$  particles (both models) with $r=0.99$
and $r=0.5$ (for the lattice model in the inhomogeneous regime). \label{APfig_dist_1d}
}
\end{figure}

In the left panel of figure~\ref{APfig_energia_1d} the decay of
kinetic energy $E(t)\equiv T(t)=\langle v^2 \rangle$ is shown. It
displays, for different values of $r$, the same behavior with two
regimes: in the HCS $E(t) \sim t^{-2}$ (Haff's law), then it decays
differently, i.e.  $E(t) \sim t^{-3/2}$. Noticeably all the curves
seem to collapse in the second regime, i.e. the asymptotic regime is
equivalent (not only in the exponent, but also in the coefficients)
for every value of $r$.

The velocity pdf's are shown in figure~\ref{APfig_dist_1d}: in the HCS
it appears the peculiar ``two-peaks'' form, which is also predicted by
the study of the Boltzmann equation in the quasi-elastic
limit~\cite{benedetto97}. In the second regime (which cannot be
described by the Boltzmann equation, because of the presence of
correlations) the pdf becomes a Gaussian.

\begin{figure}[tbp]
\centerline {\includegraphics[clip=true,width=.7\textwidth,keepaspectratio]{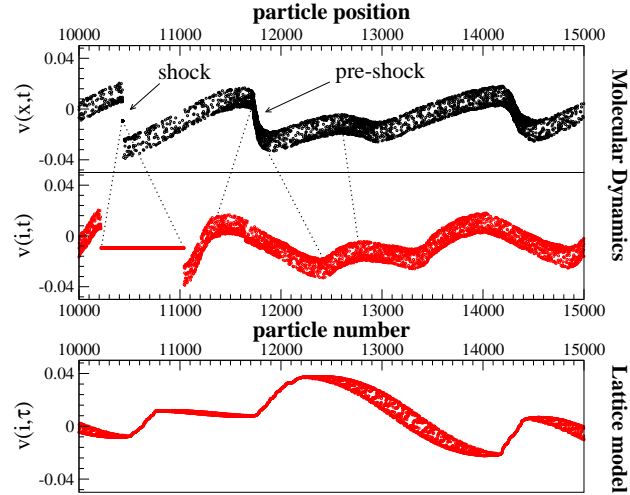}}
\caption
{Portions of the instantaneous velocity profiles for the 1D MD
  (top $v(x,t)$, middle $v(i,t)$) and for the 1D lattice
  gas model (bottom, $v(i,\tau)$). In the middle frame we display the
  MD profile against the particle label in order to
  compare the shocks and preshocks structures with the lattice gas
  model (the dotted lines show how shocks and preshocks transform
  in the two representations for the MD). Data refers to $N=2\cdot
  10^4$ particles, $r=0.99$ (both models). 
\label{APfig_shocks}  }
\end{figure}

\begin{figure}[tbp]
\centerline {\includegraphics[clip=true,width=.7\textwidth,keepaspectratio]{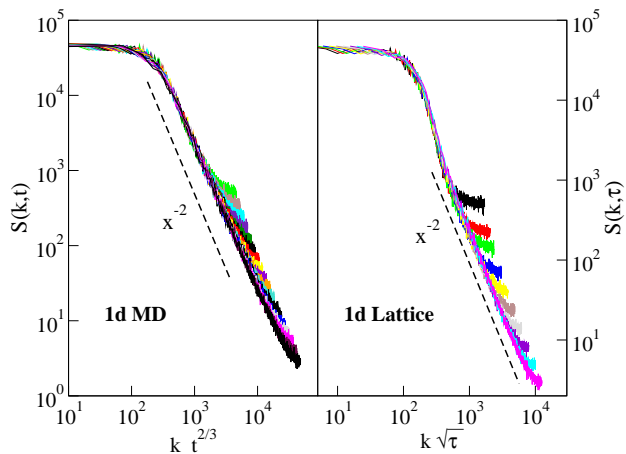}}
\caption{Structure factors $S(k,t)$ against $kt^{2/3}$ for the 1D MD 
 and against $k\tau^{1/2}$ for the 1D lattice gas model, in
 the inhomogeneous phase. Times are chosen so that the two
  systems have the same energies. Data refers to system with more than
 $N=10^5$ particles, $r=0.5$ (both models)\label{APfig_sf_1d}}
\end{figure}

What can be said about correlations in the second regime? An
inspection of the velocity profile $v_i$ vs. $x_i$ (see top panel of
figure~\ref{APfig_shocks}) suggests the presence of shocks,
i.e. asymmetric ``discontinuities'' of the velocity profile in
correspondence with peaks of density (clusters). A more quantitative
information about the velocity field is given by the study of the
structure function $S(k,t)=\hat{v}(k,t)\hat{v}(-k,t)$ where
$\hat{v}(k,t)$ is the Fourier transform of the field $v(i,t)$,
presented in the left frame of figure~\ref{APfig_sf_1d}. A good
collapse is obtained if $S(k,t)$ is plotted at different times $t$
versus $kt^{2/3} \sim k/E(t)$. The presence of short scale
defects (shocks) is put in evidence by the form of structure factor at
high values of $k$, where $S(k,t) \sim k^{-2}$: this tail of the
structure function is expected, in $d=1$, in the coarsening of a
two-phases system, due to the non-analyticity (at short distances) of
the spatial correlation function. This is also known as Porod law in
the theory of phase ordering kinetics ~\cite{porod51,porod52,bray94b}.

\subsection{The Inelastic Lattice Maxwell Model}

The Inelastic Lattice Maxwell Model in $d=1$ is defined as follows. A
set of $N$ scalar velocities are placed on $N$ sites on a ring. After
a time $\Delta \tau=2/N$ a couple of {\em neighbors} is chosen and is
updated with the usual inelastic collision rule. The only constraint
in the choice of the couple is that $v_i>v_{i+1}$: this is called the
{\em kinematic constraint} and represents the physical condition
necessary for a collision. It must be noted that the choice of
colliding particles does not depend on the modulus of the relative
velocity: this is why we consider this a ``Inelastic Maxwell Model''
embedded in a $1d$ lattice.  In this model the only measure of time is
$\tau$, i.e. the cumulated number of collisions per particle. This
makes many measures difficult to be compared with analogous MD
measurements: in fact in MD simulations $\tau(t)$ depends upon the
choice of $\delta$ (the elastic cut-off), while all observables depend
on $t$ independently of $\delta$.

The analysis of the ILMM follows closely that of MD simulations and
the results~\cite{baldassarri_epl,baldassarri_trieste} are presented,
in parallel, in the same
figures~\ref{APfig_energia_1d},~\ref{APfig_dist_1d},~\ref{APfig_shocks},~\ref{APfig_sf_1d}.

The study of the decay of energy (figure~\ref{APfig_energia_1d})
indicates that the model well reproduces the Haff regime ($E(\tau)
\sim \exp(-\gamma_0\tau)$) and displays a decay in the second regime of the form $\sim
\tau^{-1/2}$. 

The velocity pdf's reproduce exactly the ones measured in MD
simulations (see fig.~\ref{APfig_dist_1d}), showing the ``two-peaks''
form in the Haff stage and the Gaussian in the late stage. It must be
stressed again that these features are obtained simply applying to the
Inelastic Maxwell Model (which displays tails $P(v) \sim v^{-4}$) a
two-neighbors (instead of $N$-neighbors) topology.

The inspection of velocity profiles must be carefully carried: in this
model the particles cannot move and therefore a comparison with MD
should be made considering $v_i$ vs. $i$ instead of $x_i$. The good
agreement between MD and ILMM can be appreciated in the two bottom
panels of figure~\ref{APfig_shocks}: with this choice of abscissa, the
shocks observed in MD simulations appear reversed and smoothed, very
similar to the ones observed in the ILMM. Analogously the structure
factor presents similar features with MD, mainly a $k^{-2}$ behavior
at high values of $k$ (short scales) which is a signature of
topological defects. In the lattice case the collapse can be obtained
plotting $S(k,\tau)$ against $k\tau^{1/2}$, i.e.  $\sim k/E(\tau)$ as
in the MD case. We have also verified that, removing the kinematic
constraint, this structure disappears and the model behaves as
governed by a simple diffusion equation.

\section{The two-dimensional gas}

The number of interesting and realistic results obtained with the one
dimensional Inelastic Lattice Maxwell Model, in spite of its extreme
simplicity, convinced us to extend our study to two dimensions.

However, a smaller number of results is known from Molecular Dynamics
in $d=2$, especially in the correlated regime, which requires very
long simulations.  Therefore the comparison between MD and the ILMM to
the case $d=2$ is more difficult. Only recently MD simulations start
to give reliable measurements even for long times, and, where
possible, the comparison with our ILMM is promising.

More speculatively, we tried a series of less standard measurements
with two different aims: the first is to try to understand what kind
of effective equation could describe the velocity field of the ILMM;
the second is to discuss some basic hypothesis necessary to a
hydrodynamic description of this granular fluid.  After a brief review
of the known results from MD, we present the results obtained with our
2D ILMM. First the same analysis of energy decay, velocity pdf and
structure factors is carried on, as in the 1d case.  Moreover, to
better capture the nature of the topological defects generated by the
dynamics, we consider the pdf of the velocity increments.  Then we
present an analysis of temporal correlations, mainly aimed to identify
an effective equation for the velocity field.  Finally we suggest a
measurement of a hydrodynamic temperature  and we sketch a
discussion about the very presence of scale separation.

\subsection{Known results from Molecular Dynamics}

To our knowledge few MD simulations have inspected the behavior of
inelastic hard disks in the late stage, i.e. after the end of Haff
regime and the growth of instabilities. The departure from the
homogeneity has been studied by Ernst and
co-workers~\cite{noije97,noije98,noije00} and their MD simulations
have shown results compatible with their predicted $\tau^{1/2}$
scaling law for the linear size of structures (vortices and
clusters). A figure in a work of Huthmann et al.~\cite{huthmann00} was
the first suggestion that the global velocity pdf asymptotically
returns to a Gaussian (while it has fatter tails in the HCS) and
recently different works on large MD simulations have confirmed
this~\cite{nakanishi02,ben-naim02}. Recently this
``return-to-the-Gaussian'' scenario was put in strict relation with
the presence of a Burgers-like dynamics~\cite{ben-naim02}. The debate on
the asymptotic energy seems to be still open, even if there are
strong indications of a universal (independent of $r$) decay of the
form $E(t) \sim t^{-1}$~\cite{ben-naim02}. A recent work of
Isobe~\cite{isobe_private} demonstrates that many features of the
``last'' stage resemble the dynamics of 2D turbulence.

\subsection{The Inelastic Maxwell Lattice Model}

The ILMM in $d=2$ is identical to that in $d=1$ except for the fact
that the particles have a $2$-components vector velocity $\mathbf{v}_i$
and that they are placed on a triangular $2D$ lattice, i.e. every
particle has $6$ neighbors. The dynamics is the same as in $d=1$; the
kinematic constraint here reads 
$(\mathbf{v}_i-\mathbf{v}_j) \cdot (\mathbf{r}_i - \mathbf{r}_j) <0$. 

\begin{figure}[tbp]
\centerline {\includegraphics[clip=true,width=.7\textwidth,keepaspectratio]{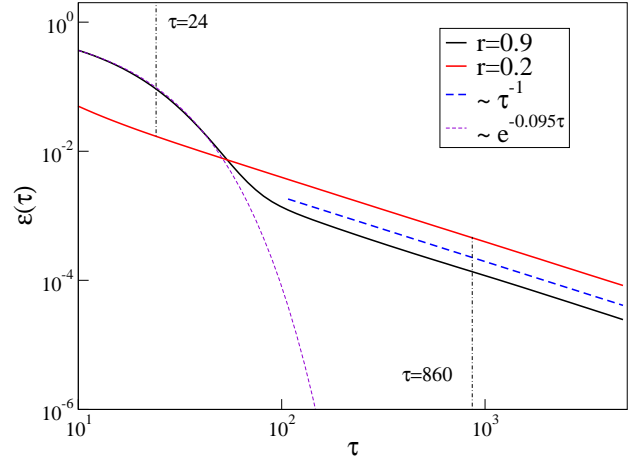}}
\caption
{Energy decay for $r=0.9$ and $r=0.2$ ($1024^2$ sites). The bold dashed
line $\sim 1/\tau$ is a guide to the eye for the asymptotic energy
decay, while the light dashed line is the exponential fit corresponding to
the Haff law $\exp(-2 \gamma_0 \tau)$. The Haff regime is too short to
be observed in the system with $r=0.2$. The two indicated times
$\tau=24$ and $\tau=860$ correspond to the plots of fig.~\ref{APfig_tempint_2d}.  \label{APfig_energia_2d}
}
\end{figure}

The decay of energy $E(\tau)$ is shown in figure~\ref{APfig_energia_2d}
for two different values of $r$. The Haff regime $E(\tau) \sim
\exp(-\gamma_0\tau)$ is well reproduced (when $r=0.2$ it terminates too early to
be appreciated). The asymptotic decay reads $E(\tau) \sim
\tau^{-1}$. Both this measurement and the decay observed in the $d=1$
case, are compatible with a universal law $E(\tau) \sim \tau^{-d/2}$
as for a purely diffusive field.

\begin{figure}[tbp]
\centerline {\includegraphics[clip=true,width=1.\textwidth,keepaspectratio]{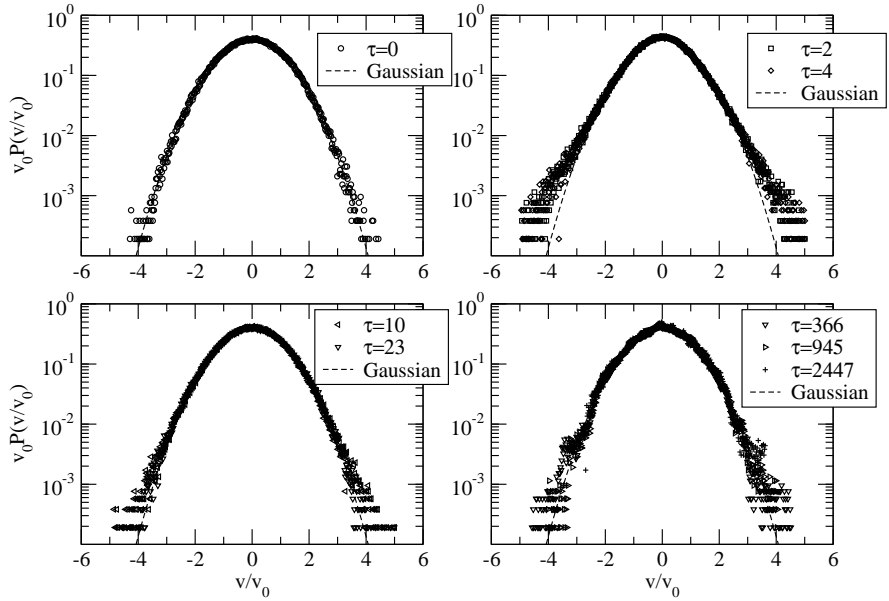}}
\caption
{Distributions of horizontal velocity at different times for the same system
with $r=0.2$ and $N=512 \times 512$. The distributions are rescaled in order
to have unit variance. The initial distribution is a Gaussian. The
distribution becomes broader in the uncorrelated phase (first regime),
and then turns back toward a Gaussian. \label{APfig_vdist_2d}
}
\end{figure}

The rescaled velocity pdf's (see figure~\ref{APfig_vdist_2d}), which are
initially Gaussian, become non-Gaussian in the Haff regime, with
larger tails, but return Gaussian in the late stage. The analysis that
follows shows that, in the late stage, the velocity field is far from
being homogeneous: the global velocity pdf is strongly influenced by
this homogeneity and its Gaussian form could be simply the signature
of the presence of many independent domains (i.e. it is just the
distribution of the average velocities of these domains).

\begin{figure}[tbp]
\centerline {\includegraphics[clip=true,height=.7\textwidth,keepaspectratio,angle=-90]{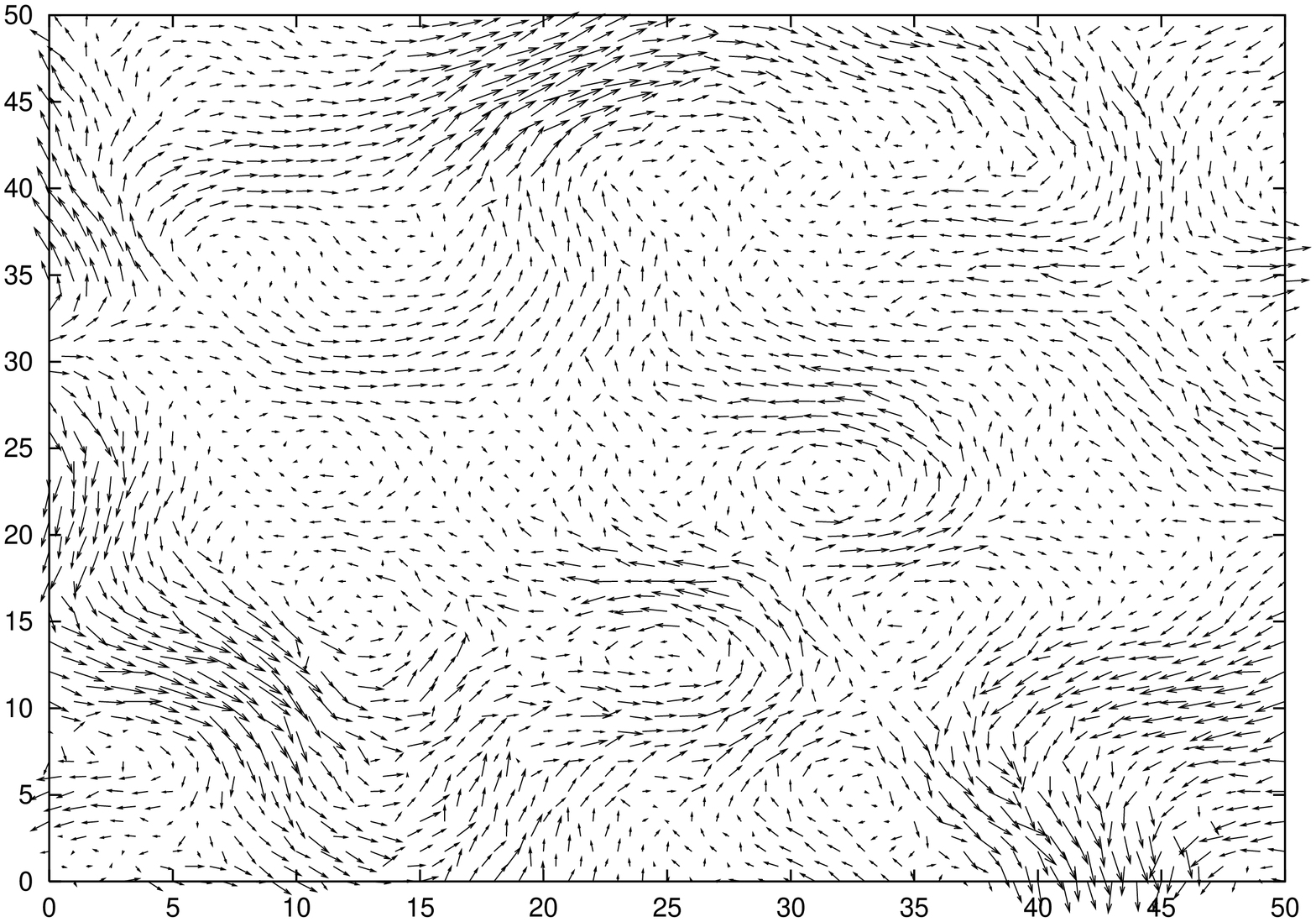}}
\caption
{ A (zoomed) snapshot of the velocity field at time $\tau=52$ for the
Inelastic Lattice Gas, $d=2$, with $r=0.7$ and size $N=512 \times 512$. The
time has been chosen at the beginning of the correlated regime. It is
evident the presence of vortices. All the velocities have been
rescaled to arbitrary units, in order to be visible.\label{APfig_vorticiA_2d}
}
\end{figure}

\begin{figure}[tbp]
\centerline {\includegraphics[clip=true,height=.7\textwidth,keepaspectratio,angle=-90]{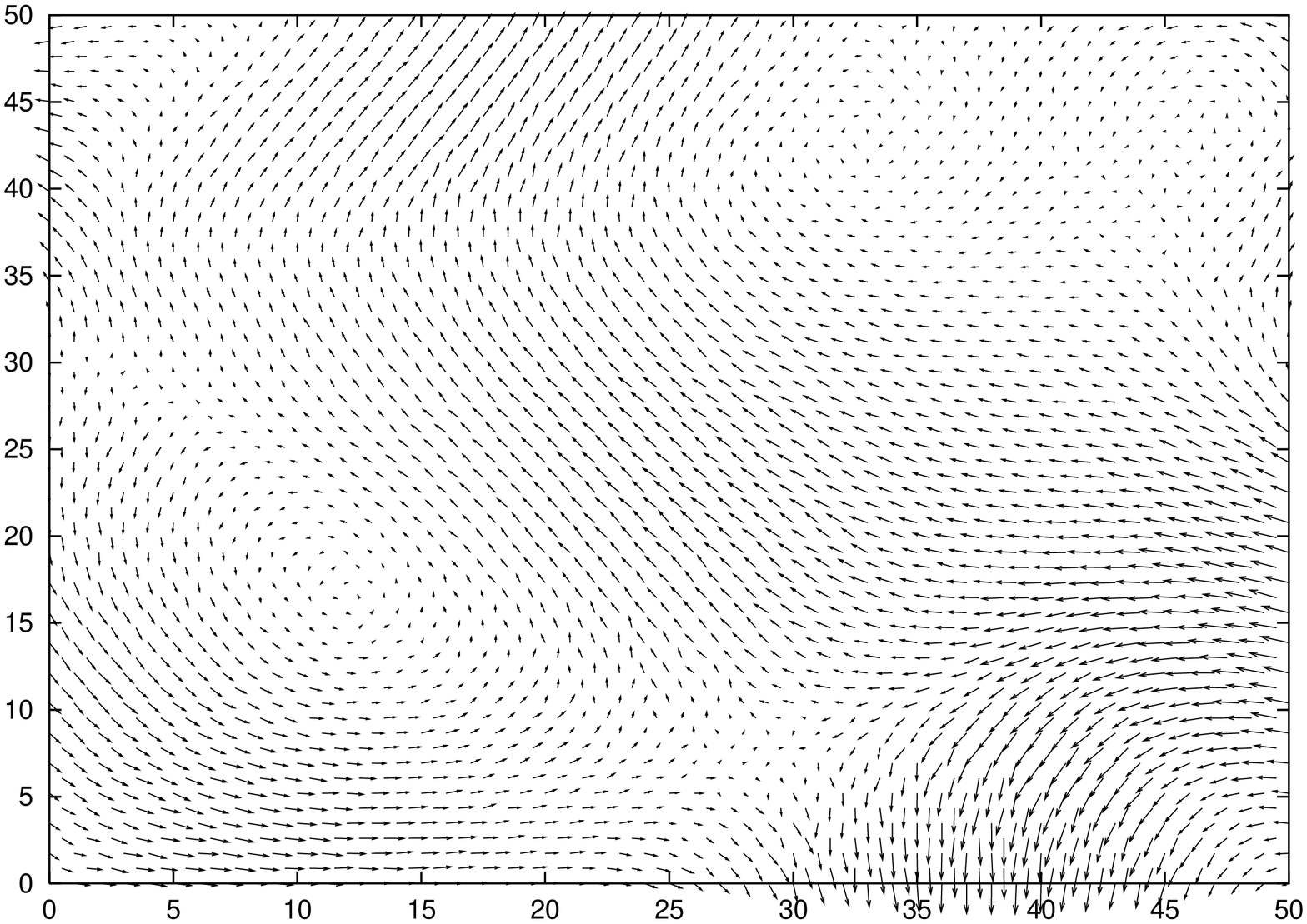}}
\caption
{Another snapshot of the same system of 
fig.~\ref{APfig_vorticiA_2d} but at a later time $\tau=535$. The diameter
of the vortices has grown up. All the velocities have been rescaled to
arbitrary units, in order to be visible.\label{APfig_vorticiB_2d}
}
\end{figure}

In the figures~\ref{APfig_vorticiA_2d} and~\ref{APfig_vorticiB_2d} we show
the presence of vortices in the velocity fields. The second figure
represents the same system at a later time, demonstrating the
coarsening of vortices (their number reduces and their size
increases). 

\begin{figure}[tbp]
\centerline {\includegraphics[clip=true,width=.7\textwidth,keepaspectratio]{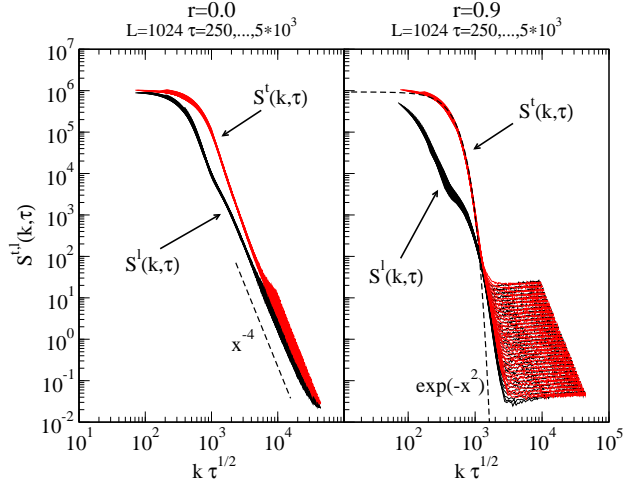}}
\caption
{ 
Data collapse of the transverse ($S^t$) and longitudinal ($S^l$)
structure functions for $r=0.$ and $r=0.9$ (system size
$1024^2$ sites, times ranging from $\tau=500$ to $\tau=10^4$). The
wave-number $k$ has been multiplied by $\sqrt{\tau}$. Notice the presence
of the plateaux for the more elastic system. For comparison we have
drawn the laws $x^{-4}$ and $\exp(-x^2)$. \label{APfig_sf_2d}
}
\end{figure}

In $d=2$ the structure function is defined as

\begin{equation}
S^{t,l}(k,t)=\sum_{\hat k} {\bf v^{t,l}}({\bf k},t){\bf v^{t,l}}(-{\bf k},t)
\end{equation}
where the superscripts $t,l$ indicate the transverse and longitudinal
components of the field with respect to the wave vector ${\bf k}$ and
the sum $\sum_{\hat k}$ is over a circular shell of radius $k$.  The
analysis of the structure factor, shown in figure~\ref{APfig_sf_2d},
is performed for two systems at different inelasticities. The common
feature is a good collapse of different time curves if plotted against
$k\tau^{1/2}$ which is the signature of a domain growth (diameter of
the vortices) $L(\tau) \sim \tau^{1/2}$. The elasticity however
changes dramatically the other features: the more elastic system has a
Gaussian structure factor at large scale (low values of $k$) and a
plateau at small scales, i.e. quasi-elastic collisions randomize
neighbour velocities while the effect of inelasticity induces a
structure at large scale. The less elastic system does not present
such a plateau, but a decay $k^{-4}$ at small scales which is
analogous to the Porod law~\cite{porod51,porod52} expected from the
presence of defects in a phase ordering process. The difference
between transversal and longitudinal component (more appreciable in
the quasi-elastic system) is coherent with the theoretical analysis
discussed in the introduction, which predicts an earlier and stronger
correlation for the transversal compoenent.

\begin{figure}[tbp]
\centerline {\includegraphics[clip=true,width=.7\textwidth,keepaspectratio]{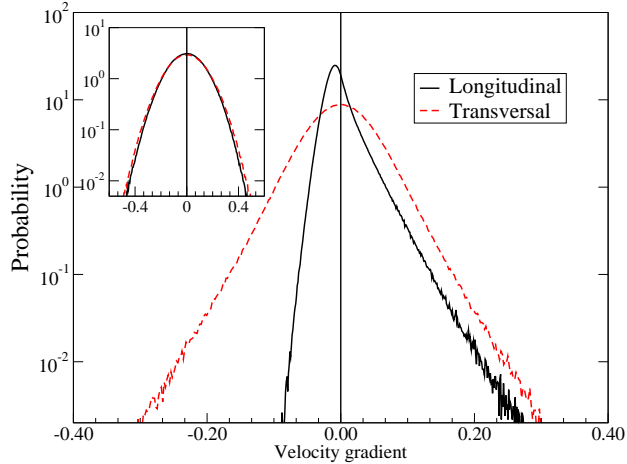}}
\caption
{ 
Probability densities of the longitudinal and transverse velocity
increments. The main figure shows the p.d.f. of the velocity differences
for $R=1$. The inset shows the Gaussian shape measured for $R=40$
(larger than $L(t)$ for this simulation: $r=0.2$, $t=620$, system
size $2048^2$).  \label{APfig_vgrad_2d}
}
\end{figure}

In the $d=2$ ILMM we have also observed shock-like phenomena. The
inspection of the distribution of longitudinal and transversal
velocity differences, defined as

\begin{subequations}
\begin{align}
\Delta_l(\mathbf{R},\mathbf{i}) &= (\mathbf{v}_{i+ R} - \mathbf{v}_i) \cdot \frac {\mathbf {R}}{R}\\
\Delta_t(\mathbf{R},\mathbf{i}) &= (\mathbf{v}_{i+ R} - \mathbf{v}_i) \times \frac {\mathbf {R}}{R}
\end{align}
\end{subequations}
is contained in figure~\ref{APfig_vgrad_2d}. The main graph shows the
case $R=1$, i.e. the distributions of gradients. Both the longitudinal
and transverval components are distributed with non-Gaussian tails,
but the longitudinal component presents also a strong asymmetry: this
means that preferentially there are strong velocity differences in the direction
of growing longitudinal velocity. This is compatible with the analysis
of shocks in the bottom panels of figure~\ref{APfig_shocks} in the $d=1$
case. When $R \gg 1$ the distribution of velocity differences is a
Gaussian.

\begin{figure}[tbp]
\centerline {\includegraphics[clip=true,width=.7\textwidth,keepaspectratio]{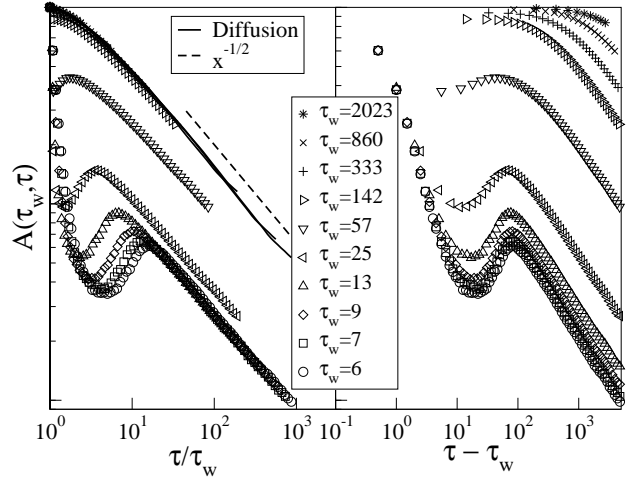}}
\caption
{ 
Angular auto-correlation function $A(\tau,\tau_w)$ for different values of
the waiting time $\tau_w$ and $r=0.9$ ($1024^2$ sites). The graph on
the left shows the convergence to the $\tau/\tau_w$ diffusive scaling
regime, for large $\tau_w$. For small $\tau_w$, a local minimum is visible
(for such a quasi elastic dynamics). In the graph on the right the
same data are plotted vs $\tau-\tau_w$: note that the small $\tau_w$ curves
tend to collapse. For higher $\tau_w$ the position of the local minimum
does not move sensibly, but its value grows and goes to $1$ for large
$\tau_w$ \label{APfig_ac_2d}
}
\end{figure}

To better study the dynamics of the system we have calculated the two-times
self-correlation of the velocity components

\begin{equation}
C(\tau_1,\tau_2)=\frac{\sum_i v_i(\tau_1) v_i(\tau_2)}{N}
\end{equation}
During a short time transient, the self-correlation function of our
model depends on $\tau_1-\tau_2$, i.e. it is time translational
invariant ({\em TTI}).  Later, $C(\tau_1,\tau_2)$ reaches an ``aging''
regime and depends only on the ratio $x=\tau_1/\tau_2$.  This {\em
TTI} transient regime is similar to what occurs during the coarsening
process of a quenched magnetic system: the self-correlation of the
local magnetization $a(\tau_w,\tau_w+\tau)$ for $\tau \ll \tau_w$
shows a {\em TTI} decay toward a constant value $m_{eq}^2(T_{quench})$
that is the square of the equilibrium magnetization, indicating that
the local magnetization is evolving in an ergodic-like
fashion. Thereafter the self-correlation decays with the {\em aging}
scaling law indicated above. Obviously, when $T_{quench}\to 0$, the
{\em TTI} transient regime disappears.  In our model the behavior of
the self-correlation is even more subtle, as the cooling process
imposes a (slowly) decreasing ``equilibrium'' temperature
$T_{quench}\to 0$: this progressively erodes the {\em TTI} regime and
better resembles a finite rate quench.  The same dependence on the
{\em TTI} manifests itself in the angular auto-correlation, shown in
figure~\ref{APfig_ac_2d}:

\begin{equation}
A(\tau,\tau_w)=\frac{1}{N}\sum_i \cos(\theta_i(\tau_w+\tau)-\theta_i(\tau_w)).
\end{equation}

The non-monotonic behavior of $A(t,t_w)$ suggests that the initial
direction of the velocity induces a change in the velocities of the
surrounding particles, which in turn generates, through a sequence of
correlated collisions, a kind of retarded field oriented as the
initial velocity. As $t_w$ increases the maximum is less and less
pronounced.

The behavior of the energy decay ($E(\tau) \sim \tau^{-d/2}$), the law
of growth of domains ($L(\tau) \sim \tau^{1/2}$) and the ``aging''
form of the two-times self-correlations, as well as the Gaussian
velocity pdf's and the Gaussian form at large scale in the structure
factors, are strong evidence that the model behaves similarly to a
system governed by the diffusion equation:

\begin{equation}
\label{APdiffusion}
\frac{\partial \phi_i}{\partial \tau} =D \sum_j \frac{\partial_i^2 \phi}{\partial r_j^2}
\end{equation}
where $\boldsymbol{\phi}$ is a suitable field depending on the
velocity field (it could be a scalar, for example the $z$-component of
the vorticity field $\boldsymbol{\nabla} \times \mathbf{v}$).

\begin{figure}[tbp]
\centerline {\includegraphics[clip=true,width=.7\textwidth,keepaspectratio]{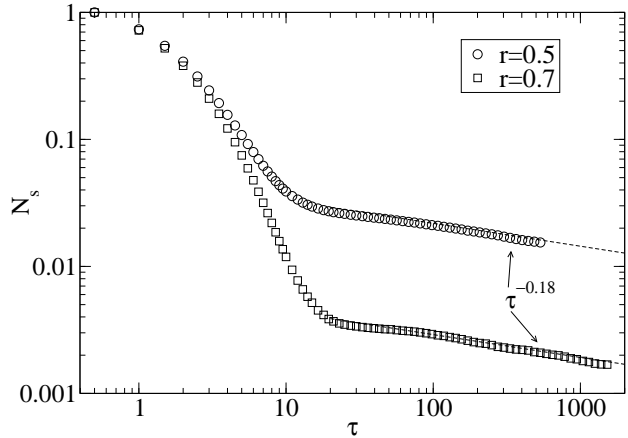}}
\caption
{ Measure of the persistence of the Inelastic Lattice Gas in $d=2$,
for two different values of the restitution coefficient. The power-law
regime corresponds to the correlated regime. The measured exponent
$\theta=0.18$ is, with this precision, equal to the persistence
exponent numerically measured in the diffusive
dynamics.\label{APfig_pers_2d} }
\end{figure}

To  perform a further check of  this conjecture,  we have measured the
persistence   exponent~\cite{bray94}  of   the model,  which   is very
sensitive to  the class  of universality. We  have counted  the number
$N_s(\tau)$  of sites where  the $x$  velocity component never changed
sign from the starting time of the dynamics up to  time $\tau$. In the
correlated   regime we observe    $N_s(\tau) \sim \tau^{-\theta}$ with
$\theta=0.18$ which is equal, up to this precision, to the persistence
exponent numerically measured for the diffusive dynamics.

\begin{figure}[tbp]
\centerline{\includegraphics[clip=true,width=.7\textwidth,keepaspectratio]{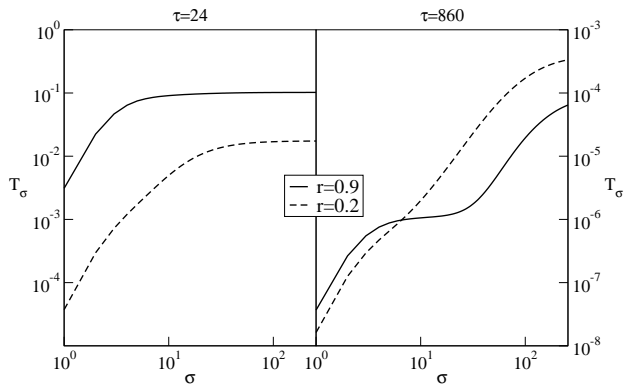}}
\caption
{The scale dependent temperature, $T_{\sigma}$, defined as function of
the coarse graining size $\sigma$ for $\tau=24$ (incoherent regime)
and $\tau=860$ (correlated regime for both choices of $r$): the system
is the same of figure~\ref{APfig_energia_2d}.  In the correlated
regime the more elastic case presents a plateau at intermediate
wavelengths, indicating a well defined {\em mesoscopic} temperature,
and therefore a clear separation between the microscopic and
the macroscopic scales.\label{APfig_tempint_2d} }
\end{figure}

In the spirit of verifying the possibility of a hydrodynamic
(mesoscopic) description of this model, we have also analysed the 
average local granular temperature $T_{\sigma}$, defined as:

\begin{equation}
T_{\sigma}=\langle |\mathbf{v}- \langle \mathbf{v} \rangle_\sigma |^2 \rangle_\sigma
\end{equation}
where $\langle ... \rangle_\sigma$ means an average on a region of linear size $\sigma$.

If we call $L(t)$ a characteristic correlation length of the system,
since when $\sigma \gg L(t)$ the local average tends to the global
(zero) momentum, then $\lim_{\sigma \to \infty} T_{\sigma} = E$.  For
$\sigma < L(t)$, instead, $T_{\sigma} < E $. The behavior of
$T_\sigma$ in the uncorrelated (Haff) regime and in the correlated
(asymptotic) regime for two different values of $r$ is presented in
figure~\ref{APfig_tempint_2d}. For quasi elastic systems $T_\sigma$
exhibits a plateau for $1\ll \sigma\ll L(t)$ that identifies the
strength of the internal noise (see also the plateau in the structure
factor, figure~\ref{APfig_sf_2d}) and indicates the mesoscopic scale
necessary for a hydrodynamics description.  The local temperature
ceases to be well defined for smaller $r$, revealing an absence of
scale separation between microscopic and macroscopic fluctuations in
the strongly inelastic regime~\cite{goldhirsch99}.

\section{Conclusions}

We have analysed essentially three different problems with a strong
interconnection, in the framework of the physics of cooling granular
gases. Our first step (IMM) was to demonstrate that a simplified 1D
version of the Boltzmann equation for inelastic gases can be exactly
solved by a $P(v)$ with power law tails. However, realistic granular
systems never show such large tails: the key feature that must be
added to this oversimplified model is a realistic topology. Therefore
our second step (ILMM, $d=1$) was to embed the IMM on a $1D$
lattice. The $P(v)$ changes dramatically and becomes identical to
those measured in MD simulations, also in the velocity-correlated
stage. Moreover, the ILMM allows for a study of spatial correlations,
resulting in shock-like structures very similar to the one observed in
MD and strong analogies with the MD structure functions. Our last step
(ILMM, $d=2$) has been the passage from $d=1$ to $d=2$, which mainly
results in the appearance of vortices. Vortices are ubiquitous in
granular gases experiments and simulations. The ILMM simply produces
vortices as a result of the competion between the parallelizing effect
of inelastic collisions and the constraint given by the conservation
of total momentum. Moreover, a large set of evidences (energy decay,
structure factors, Gaussian velocity pdf's, aging self-correlations,
persistence exponent) indicate that the dynamical behavior of the
system is compatible with a diffusion equation, even if short scale
defects (appearing as shocks, internal noise and tails {\em \`a la}
Porod in the structure functions) make this model richer.

\end{document}